%% file: root.tex
\documentclass[a4paper, 10pt, conference]{ieeeconf}      

\IEEEoverridecommandlockouts                              

\usepackage{cite}
\usepackage{balance}
\usepackage{amsmath,amssymb,amsfonts}
\usepackage{soul}
\usepackage{algorithmic}
\usepackage[hidelinks]{hyperref}
\usepackage[linesnumbered,ruled,vlined]{algorithm2e}
\usepackage{graphicx}
\usepackage{textcomp}
\usepackage{xcolor}
\usepackage{caption}
\usepackage{subcaption}
\usepackage{multirow}
\usepackage{balance}

\usepackage{array}
\newcommand{\PreserveBackslash}[1]{\let\temp=\\#1\let\\=\temp}
\newcolumntype{C}[1]{>{\PreserveBackslash\centering}p{#1}}

\title{\LARGE \bf
Ant Colony Optimization for Cooperative Inspection Path Planning Using Multiple Unmanned Aerial Vehicles
}

\author{Duy Nam Bui$^{1}$, Thuy Ngan Duong$^{2}$, Manh Duong Phung$^{3}$
\thanks{$^{1}$Duy Nam Bui is with the Vietnam National University, Hanoi, Vietnam. {e-mail: \tt\footnotesize duynam@ieee.org}}
\thanks{$^{2}$Thuy Ngan Duong is with the Ulsan National Institute of Science and Technology, Ulsan, Korea. {e-mail: \tt\footnotesize nganduong@unist.ac.kr}}
\thanks{$^{3}$Manh Duong Phung is with the  Fulbright University Vietnam, Ho Chi Minh City, Vietnam. {e-mail: \tt\footnotesize duong.phung@fulbright.edu.vn}}
}

\begin{document}

\maketitle
\thispagestyle{empty}
\pagestyle{empty}

\begin{abstract}
This paper presents a new swarm intelligence-based approach to deal with the cooperative path planning problem of unmanned aerial vehicles (UAVs), which is essential for the automatic inspection of infrastructure. The approach uses a 3D model of the structure to generate viewpoints for the UAVs. The calculation of the viewpoints considers the constraints related to the UAV formation model, camera parameters, and requirements for data post-processing. The viewpoints are then used as input to formulate the path planning as an extended traveling salesman problem and the definition of a new cost function. Ant colony optimization is finally used to solve the problem to yield optimal inspection paths. Experiments with 3D models of real structures have been conducted to evaluate the performance of the proposed approach. The results show that our system is not only capable of generating feasible inspection paths for UAVs but also reducing the path length by 29.47\% for complex structures when compared with another heuristic approach. The source code of the algorithm can be found at {\fontfamily{pcr}\selectfont \url{https://github.com/duynamrcv/aco\_3d\_ipp}}.

\end{abstract}

\input{1_introduction.tex}
\input{2_problem.tex}
\input{3_method.tex}
\input{4_result.tex}

\section{Conclusion} \label{sec:con}
In this work, we have presented a new approach for the cooperative path planning problem to inspect 3D structures using a group of UAVs. The approach combines our calculation of the viewpoints for UAV formation to collect visual data of the structure and an optimal path planning method using nature-inspired intelligence named ACO. The evaluation results show that the proposed approach is superior to other heuristic algorithms in generating inspection paths for complex 3D structures. The results thus confirm the efficiency and applicability of the proposed method for real-world scenarios.
\section*{Acknowledgement}
Duy Nam Bui was funded by the Master, PhD Scholarship Programme of Vingroup Innovation Foundation (VINIF), code VINIF.2022.Ths.057.
\balance
\bibliographystyle{IEEEtran}  
\bibliography{ref} 

\end{document}

%% file: 1_introduction.tex
\section{Introduction}
Unmanned aerial vehicles (UAVs) with their high flexibility and versatile deployment play an important role in structural health monitoring (SHM) to carry out non-destructive inspection, evaluation of cracks or damages, and prediction of catastrophic failures \cite{PHUNG201725,8593930,WANG202213}. One of the essential tasks in SHM is solving the inspection path planning (IPP) problem as it generates the path the UAVs need to follow to collect information about the structure \cite{PHUNG201725, WANG202213}. However, there are few IPP methods for aerial robotics, particularly when it comes to collaborative inspection (C-IPP). The C-IPP involves multiple UAVs cooperating in a formation to enhance inspection solutions, such as decreasing the total inspection time \cite{Wu2021, IVIC2023104709}. 

In the literature, most works address the inspection problem in 2D space \cite{Wu2021,PHUNG201725}. For 3D structures, most studies deal with it by using a visual approach that relies on image sensors equipped on UAVs \cite{PHUNG201725,LI201483,IVIC2023104709}.  In \cite{IVIC2023104709}, 3D visual inspection trajectories for UAVs are generated based on artificial potential field to inspect complex structures. 
In \cite{MANSOURI2018118}, a heuristic method that splits the inspecting structure into multiple layers is used to generate paths for a group of UAVs to carry out the visual inspection. While these methods create feasible flight paths for UAVs, the optimality of those paths is hardly addressed.

In another direction, the growing use of artificial intelligence is transforming the way the operation and management of SHM tasks are conducted \cite{doi:10.1080/17508975.2019.1613219}. The major topics that make significant contributions include expert systems, fuzzy logic, machine learning, and swarm intelligence. In particular, the application of swarm intelligence (SI) algorithms in the SHM problem is being widely studied \cite{PHUNG201725,9341089,SHANG2020113535}. SI algorithms are inspired by the collective behaviors of decentralized, self-organizing, natural, or man-made systems. The popular algorithms that can be referred to include the genetic algorithm (GA), particle swarm optimization (PSO) and ant colony optimization (ACO). In \cite{9341089}, a modified genetic algorithm was developed to inspect large-scale, complex 3D structures by using multiple UAVs. In \cite{SHANG2020113535}, PSO is used to address the co-optimal coverage path planning problem, where it concurrently optimizes the path of the UAV, the quality of the captured images, and the computational complexity. The nearest neighbor combined with GA is used in \cite{Nagasawa2021} to optimize the coverage paths for multiple UAVs to reconstruct the 3D model of post-disaster damaged buildings. However, most studies only consider solving the inspection path for individual UAVs or discretizing the inspection space to simplify the IPP problem. The C-IPP for multiple UAVs is therefore still a challenging problem that needs further investigation.

This paper presents a new approach to the C-IPP problem for multiple UAVs to carry out inspection tasks. The approach utilizes an effective sampling-based method to calculate the viewpoint positions. The overall problem is then modeled as an extended traveling salesman problem (TSP) and solved by the ACO algorithm to obtain the optimal inspection path for a given structure. The main contributions of our work are threefold:
(i) model the UAV formation based on the camera carried on each individual UAV and then propose a method to generate viewpoints for visual data collection;
(ii) formulate the inspection path planning as an extended TSP problem with a new fitness function and solve it using the ACO algorithm;
(iii) experiment with real-world 3D models to benchmark the performance of the proposed method in comparison with other algorithms.

%% file: 2_problem.tex
\section{System overview}

The problem of collaborative inspection path planning (C-IPP) is finding efficient paths that guide autonomous systems such as unmanned aerial vehicles to inspect the region of interest with onboard sensors, e.g., camera sensors. In this paper, the use of swarm intelligence is investigated to perform the C-IPP for a group of UAVs to carry out the inspection task. The inspecting structure such as a building or a bridge is known, e.g., by using building information modeling (BIM) or 3D lidar scanners \cite{SON2015172,BOLOURIAN2020103250}. The inspection path for the UAV formation is defined as sequentially connected waypoints that indicate the location of the UAV formation. Path efficiency is measured in path length. The path also needs to satisfy constraints related to the sensor's field of view, the distance from the UAVs to the inspecting surface, and requirements for image stitching.

\begin{figure}
    \centering
    \includegraphics[width=0.45\textwidth]{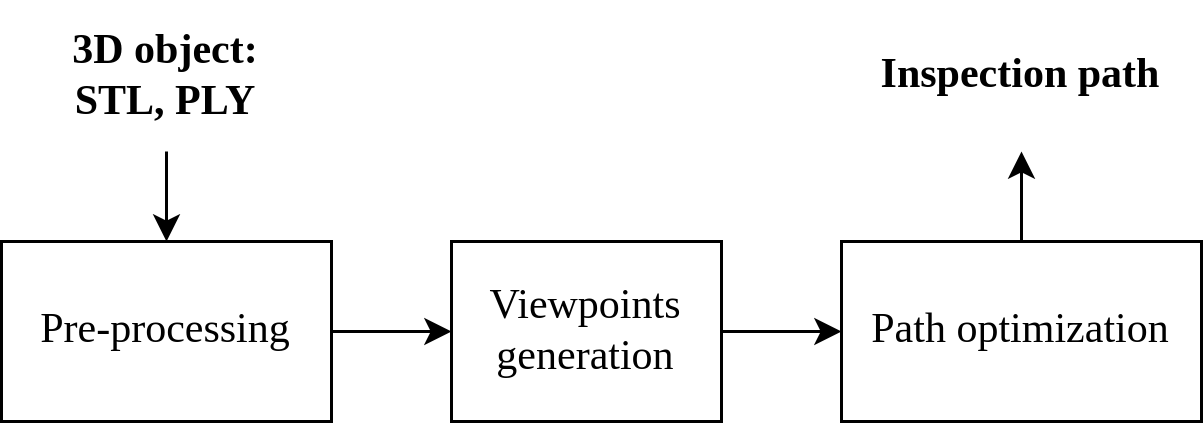}
    \caption{Overview of the proposed method }
    \label{fig:overview}
\end{figure}

Our approach to the C-IPP problem is presented in Fig.\ref{fig:overview}. The system first takes 3D models of the structure such as the STL or PLY files as the input. It then pre-processes them to sample down and filter noises before passing through the viewpoints generation module. This module calculates the points that the UAVs need to fly through to collect visual data of the structure. The viewpoints are then fed to the path optimization module that uses the ACO to generate inspection paths. 

\section{Generation of viewpoints for UAV formation} \label{sec:gen}

To collect visual information of the structure's surface, it is necessary to determine the viewpoints at which the UAVs take photos. The calculation of those viewpoints depends on the UAV formation, the camera parameters, and the requirements for post-processing with details as follows. 

\subsection{Visual-based UAV formation model}

\begin{figure}
\centering
\includegraphics[width=0.3\textwidth]{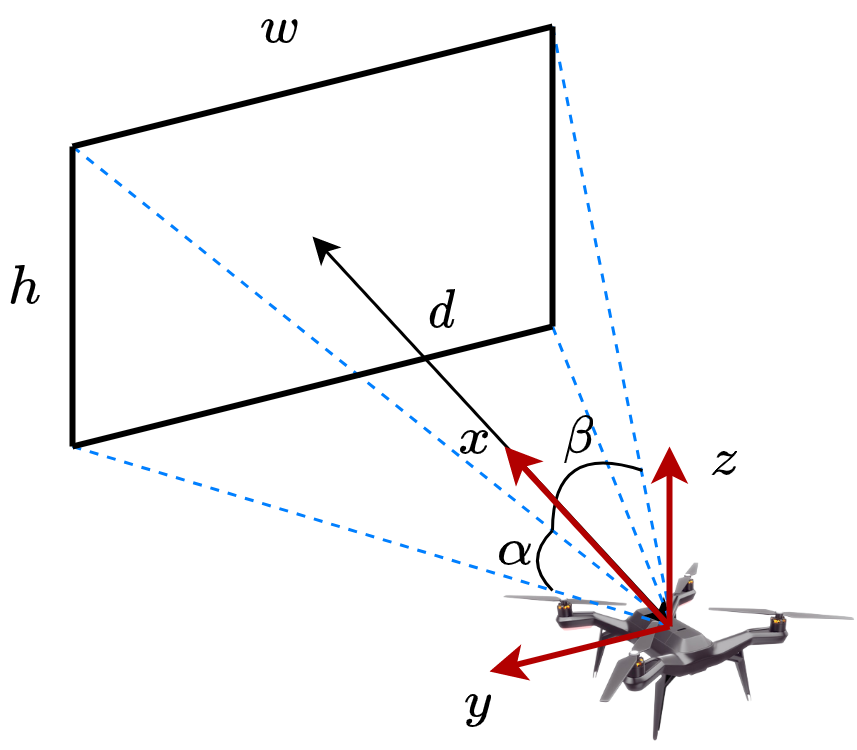}
\caption{The camera's field of view}
\label{fig:fov}
\end{figure}

The formation model used in this work is the virtual leader-follower \cite{9496675}. In this model, the leader is a non-physical UAV used as a reference for other UAVs, the followers, to determine their location. Each follower carries an RGB camera that can be stabilized and controlled via a three-axis gimbal. The gimbal controls the camera so that it is perpendicular to the inspecting surface. The footprint of the camera on the inspecting surface is illustrated as shown in Fig. \ref{fig:fov}. Let $d$ be the working distance from the UAV to the structure's surface so that the capturing photos meet the resolution required for the inspection task \cite{PHUNG201725}. Height $h$ and width $w$ of the camera's footprint are then given by:
\begin{equation}
    \begin{aligned}
        h&= 2d\tan\dfrac{\alpha}{2}\\
        w&= 2d\tan\dfrac{\beta}{2},
    \end{aligned}
    \label{eqn:camera}
\end{equation}
where $\alpha$ and $\beta$ are the horizontal and vertical angles of the camera's field of view (FOV), respectively. The overall footprint is then the union of the individual footprints but includes overlapping areas for image stitching. This area is given by $o_{ij}=h_{ij}\times w_{ij}$, $0< h_{ij}\leq h$, $0< w_{ij}\leq w$, where $h_{ij}$ and $w_{ij}$ respectively be the width and height of the overlapping area between the footprints of UAV $i$ and $j$.

\begin{figure}
\centering
\includegraphics[width=0.32\textwidth]{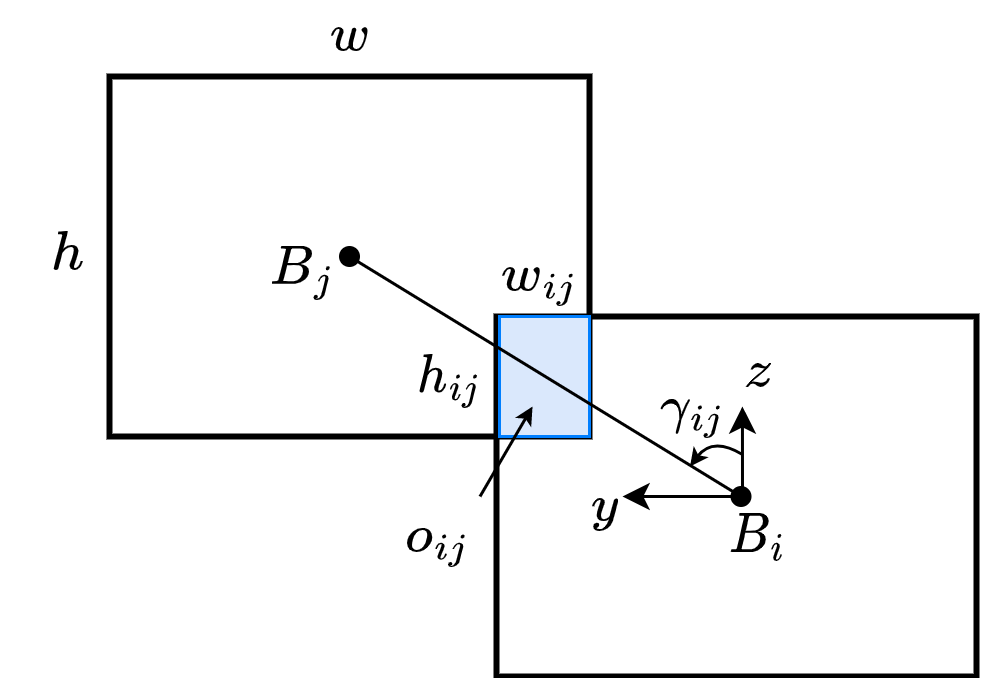}
\caption{Overlapping area of two footprints}
\label{fig:overlap}
\end{figure}

Without loss of generality, the camera points along the $x-$axis of the body frame and the local frame attached to each footprint is defined as shown in Fig. \ref{fig:overlap}, where $B_i$ is the center of footprint $i$. The distances of between UAVs $i$ and $j$, $B_iB_j$, along the $y$ and $z$ axes are given by $(w-w_{ij})$ and $(h-h_{ij})$, respectively. Denote $\gamma_{ij}$ as the angle formed by $B_iB_j$ and $B_{iz}$, and $\delta_{ij}=\overrightarrow{B_iB_j}$ as the vector representing the difference in position between UAVs $i$ and $j$. This difference is computed as follows:
\begin{equation}
    \label{eq:smallDelta}
    \delta_{ij}=\overrightarrow{B_{i}B_{j}}=\left[\begin{array}{c}
0\\
\left(w-w_{ij}\right)\text{sgn}\left(\sin\gamma_{ij}\right)\\
\left(h-h_{ij}\right)\text{sgn}\left(\cos\gamma_{ij}\right)
\end{array}\right],
\end{equation}
where $\text{sgn}\left(\cdot\right)$ denotes the sign function. Let $\Delta_i$ and $\Delta_{j}$ be the vectors from the virtual leader to UAV $i$ and UAV $j$, respectively. $\Delta_{j}$ is then computed as:
\begin{equation}
    \Delta_{j} = \Delta_i+\delta_{ij}.
    \label{eq:delta}
\end{equation}
From (\ref{eq:smallDelta}) and (\ref{eq:delta}), vectors $\Delta_i,\forall i\in\left\{1,...,n\right\}$, can be computed so that the corresponding position of each UAV satisfies the target formation. The largest width, $w_f$, and height, $h_f$, of the formation footprint are then obtained to calculate viewpoints.
\subsection{Viewpoint generation}
\begin{figure}
    \centering
    \includegraphics[width=0.48\textwidth]{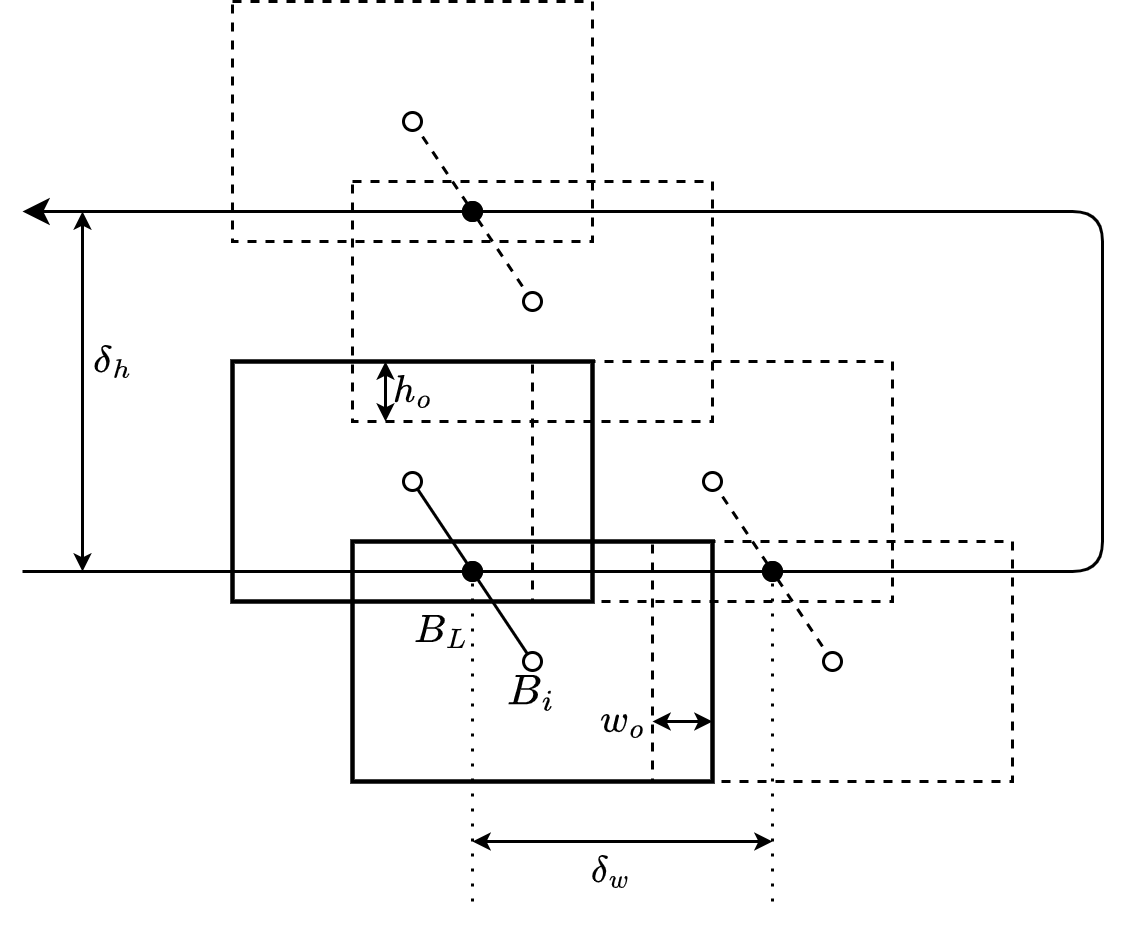}
    \caption{Horizontal and vertical overlapping requirements}
    \label{fig:requirement}
\end{figure}

\begin{figure*}
    \centering
    \includegraphics[width=\textwidth]{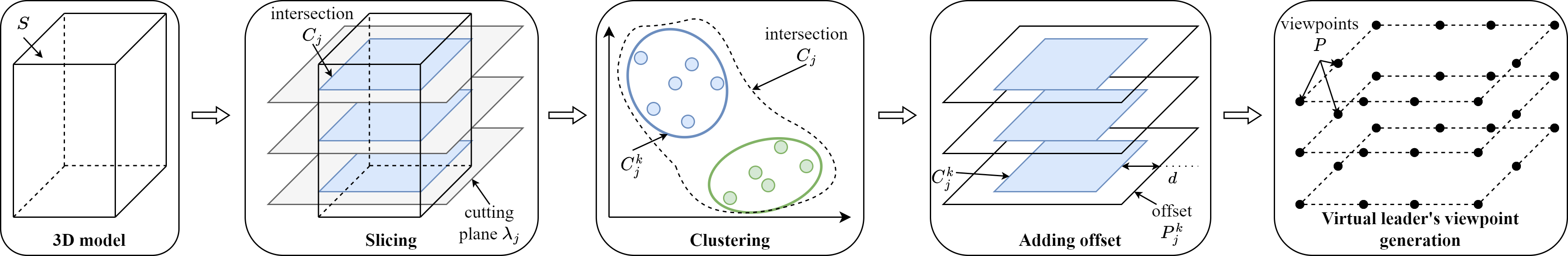}
    \caption{Virtual leader's viewpoint generation}
    \label{fig:generation}
\end{figure*}
Given the overall footprint of the formation, the viewpoints at which the UAVs take photos of the surface can be determined as illustrated in Fig. \ref{fig:requirement}. The required distances between two consecutive points in the horizontal direction, $\delta_w$, and vertical direction, $\delta_h$, are computed based on the overlapping width, $w_o$, and height, $h_o$, as:
\begin{equation}
\begin{aligned}
&\delta_w = w_f-w_o,\\
&\delta_h = h_f-h_o.
\end{aligned}
\end{equation}

The 3D model of the structure, denoted as $S$, is then sliced by multiple horizontal planes, defined as
$\lambda_j$, as shown in Fig. \ref{fig:generation}. The value of $\lambda_j$ ranges from $(\min_zS+\dfrac{h_f}{2})$ to $(\max_zS-\dfrac{h_f}{2})$ along the $z-$axis with step $\delta_h$. The intersection between the 3D model and plane $\lambda_j$ is computed as
\begin{equation}
    C_j=\left\{\left(x,y,z\right)\in\mathbb{R}^3:\left\langle \overrightarrow{v},S\left(x,y,z\right)\right\rangle -\lambda_{j}=0\right\},
\end{equation}
where $C_j\left(x,y,z\right)$ are the points of the intersecting surface, $\overrightarrow{v}=\left[0,0,1\right]^T$ is the normal vector of the plane, and $\left\langle \cdot,\cdot\right\rangle$ is the inner product operation \cite{MANSOURI2018118}. For complex structures, $C_j$ may include more than one cluster of points, and thus should be processed by clustering algorithms such as DBSCAN \cite{10.5555/3001460.3001507}, to obtain clusters $C_j^k(x,y,z)$.

Let $d$ be the distance from the UAV formation to the inspecting surface. A viewpoint $P_j^k$ is then obtained by extending the intersection set by $d$ as:
\begin{equation}
    P_{j}^k\left(x,y,z\right)=C_{j}^k\left(x,y,z\right)\pm d\overrightarrow{v}_{j},
    \label{eqn:viewpoints}
\end{equation}
where $\overrightarrow{v}_{j}$ is the normal vector of the inspecting surface pointing through $C_j^k(x,y,z)$ and $\pm$ indicates the direction of the normal vector \cite{LIU2007240}. The set of all viewpoints, $P$, is then obtained by merging all $P_j^k$ and down-sampling them by distance $\delta_w$. 

%% file: 3_method.tex
\section{Ant colony optimization for inspection path planning}\label{sec:aco}
This section presents our formulation and implementation of the ACO algorithm to generate inspection paths.

\subsection{Inspection path planning problem}
\label{sec:IPPproblem}

After determining the set of viewpoints $P$, it is necessary to plan a path through these points that the UAVs can follow. The path should be optimal and meet the constraints related to UAV operation. Since the viewpoints are generated for the formation, the planning path will be assigned to the virtual leader. Our approach is modeling this problem as an extended TSP problem, where the goal is to find a Hamiltonian path from the set of viewpoints. The TSP problem is formulated as a graph $G=(P, E, F)$, where $P =\{P_1,P_2,...,P_m\}$ is the set of viewpoints, $E = \{(i,j)|(j,i)\in p \times p\}$ is the set of edges connecting every two vertices with cost value $F_{ij}$ \cite{10.1007/978-981-13-6001-5_32}. The objective is to minimize the cost of the travel tour, $J$, which is described as follows:
\begin{equation}
    J=\sum_{i=1}^{m}\left(\sum_{j\neq i,j=1}^{m}F_{ij}\chi_{ij}\right),
    \label{eqn:J}
\end{equation}
where $\chi_{ij} = \{0,1\}$ represents whether there is a direct edge connecting two nodes $i$ and $j$.

$F_{ij}$ is the traveled cost from point $i$ to $j$ and is evaluated via a sequential convex function as follows:
\begin{equation}
    F_{ij}=w_{1}\left\Vert \overrightarrow{P_{i}^{xy}P_{j}^{xy}}\right\Vert+w_{2}\left\Vert \overrightarrow{P_{i}^{z}P_{j}^{z}}\right\Vert,
    \label{eqn:Fij}
\end{equation}
where $P_{i}=\left[P_{i}^x,P_{i}^y,P_{i}^z\right]^T$ denotes the position of the viewpoints in 3D space, $P_{i}^{xy}$ denotes the $x$ and $y$ components of $P_{i}$ and $P_{i}^{z}$ denotes the $z$ component; 
$w_1$ and $w_2$ are positive weights. We select $w_2 > w_1$ to punish more the difference in the $z$-axis since it affects the altitude of the UAV formation. 



\subsection{Optimal inspection path planning using Ant Colony Optimization}
The extended TSP problem formulated in (\ref{eqn:J}) - (\ref{eqn:Fij}) is NP-hard that cannot be solved by analytical methods. Instead, we propose to use a metaheuristic optimization technique called the ACO to take advantage of swarm intelligence in solving complex combinatorial optimization problems \cite{4129846}. This algorithm is inspired by the ants foraging in which ants lay chemical pheromone trails to discover efficient paths between their nest and food supplies. The pheromone $\tau_{ij}$ contributed by $l$ ants to edge $E_{ij}$ is computed as follows:
\begin{equation}
    \tau_{ij}=(1-\varrho)\tau_{ij}+\sum_{k=1}^{l}\Delta\tau_{ij}^{k},
    \label{eqn:pheromone}
\end{equation}
where $\varrho$ is the pheromone evaporation rate, and $\Delta\tau_{ij}^{k}$ is the pheromone quantity that ant $k$ leaves on edge $E_{ij}$. It is given by:
\begin{equation}
    \Delta\tau_{ij}^{k}=\chi_{ij}\dfrac{Q}{F_{ij}},
    \label{eqn:delta_tau}
\end{equation}
where $Q$ is a positive constant. The probability of an ant $k$ moving from state $i$ to state $j$ is computed as:
\begin{equation}
    \rho_{ij}^{k}=\dfrac{\left(\tau_{ij}\right)^{a}\left(\eta_{ij}\right)^{b}}{\sum\limits_{z\in\Omega}\left(\tau_{iz}\right)^{a}\left(\eta_{iz}\right)^{b}},
    \label{eqn:prob}
\end{equation}
where $\Omega$ is the set of unvisited viewpoints of ant $k$ at state $i$, $a$ and $b$ are the parameters representing the influence of pheromones and heuristic information, and $\eta_{ij}=(F_{ij})^{-1}$ is the heuristic information.


\begin{algorithm}
\caption{Pseudocode for the ACO algorithm}\label{alg:ACO}
    \tcc{Initialization}
    Get the list of $m$ viewpoints \;
    Initialize parameters $l$, $a$, $b$, $Q$, $\varrho$\;
    Initialize pheromone $\tau(0)$ and calculate $\eta$ matrix\;
    
    \tcc{Searching}
    \While{iter $\leq$ max\_iteration}
    {
        \For{Ant $k$ in colony}{
            \For{i $\leftarrow 1:(m-1)$ }{
                Choose the next viewpoint $j$  with the probability $\rho^k_{ij}$\tcc*[r]{Equation  \ref{eqn:prob}}
            }
            Evaluate $J_k$ of tour given by ant $k$\tcc*[r]{Equation  \ref{eqn:J}}
        }
        Update pheromone quantity $\tau_{ij}$ in each edge\tcc*[r]{Equation  \ref{eqn:pheromone}, \ref{eqn:delta_tau}}
        Store the best solution\;
    }
        Return the best solution as the inspection path\;
\end{algorithm}

The implementation of the ACO algorithm to solve the extended TSP problem for 3D inspection path planning is described in Algorithm \ref{alg:ACO}. Here, each viewpoint is considered as a node of the path and the line segment connecting two nodes is represented as an edge. Initially, the amount of pheromone $\tau_{ij}$ in each edge is set equally. Each time, an ant in the colony chooses its route based on probability $\rho^k$ computed in \eqref{eqn:prob} and then calculates the fitness of that path using \eqref{eqn:J}. After traveling through an edge $E_{ij}$, each ant $k$ leaves an amount of pheromone trail $\Delta\tau_{ij}^k$ on that edge inversely proportional to the value $F_{ij}$ of the edge. The total quantity of pheromones $\tau_{ij}$ in each edge is then updated based on the visited ants and the evaporation amount as in \eqref{eqn:pheromone} and \eqref{eqn:delta_tau}. From there, the probability matrix $\rho^k$ is updated to guide the next ants to choose better routes. The procedure is reiterated until it reaches the maximum allowable number of iterations. The path having the smallest value of $J$ is chosen as the inspection path and is assigned to the virtual leader of the UAV formation. The real follower UAVs then use that path as a reference to adjust their position according to (\ref{eq:delta}) to form the expected formation for data collection.

%% file: 4_result.tex
\section{Results} \label{sec:res}
\begin{figure}
    \centering
    \includegraphics[width=0.28\textwidth]{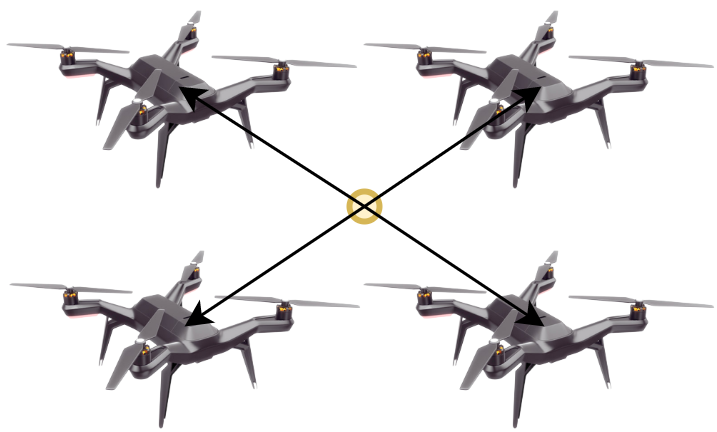}
    \caption{The inspection formation structure}
    \label{fig:topology}
\end{figure}

\begin{figure*}
     \centering
     \begin{subfigure}[b]{0.25\textwidth}
         \centering
         \includegraphics[width=\textwidth]{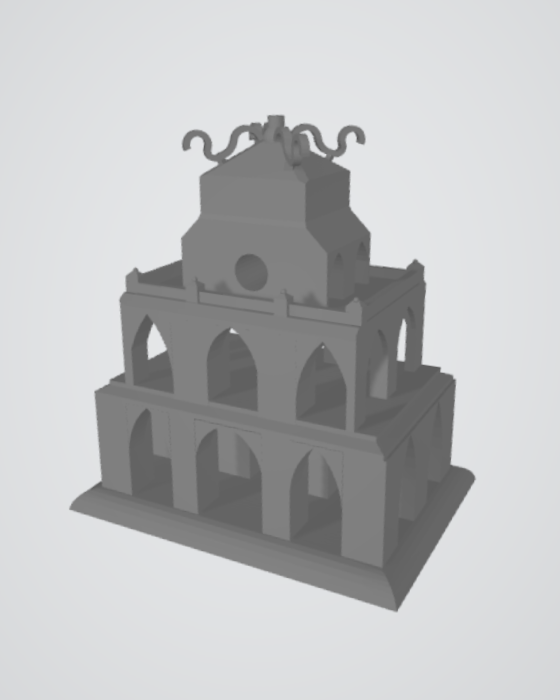}
         \caption{3D model}
         \label{fig:stl}
     \end{subfigure}
     \begin{subfigure}[b]{0.35\textwidth}
         \centering
         \includegraphics[width=\textwidth]{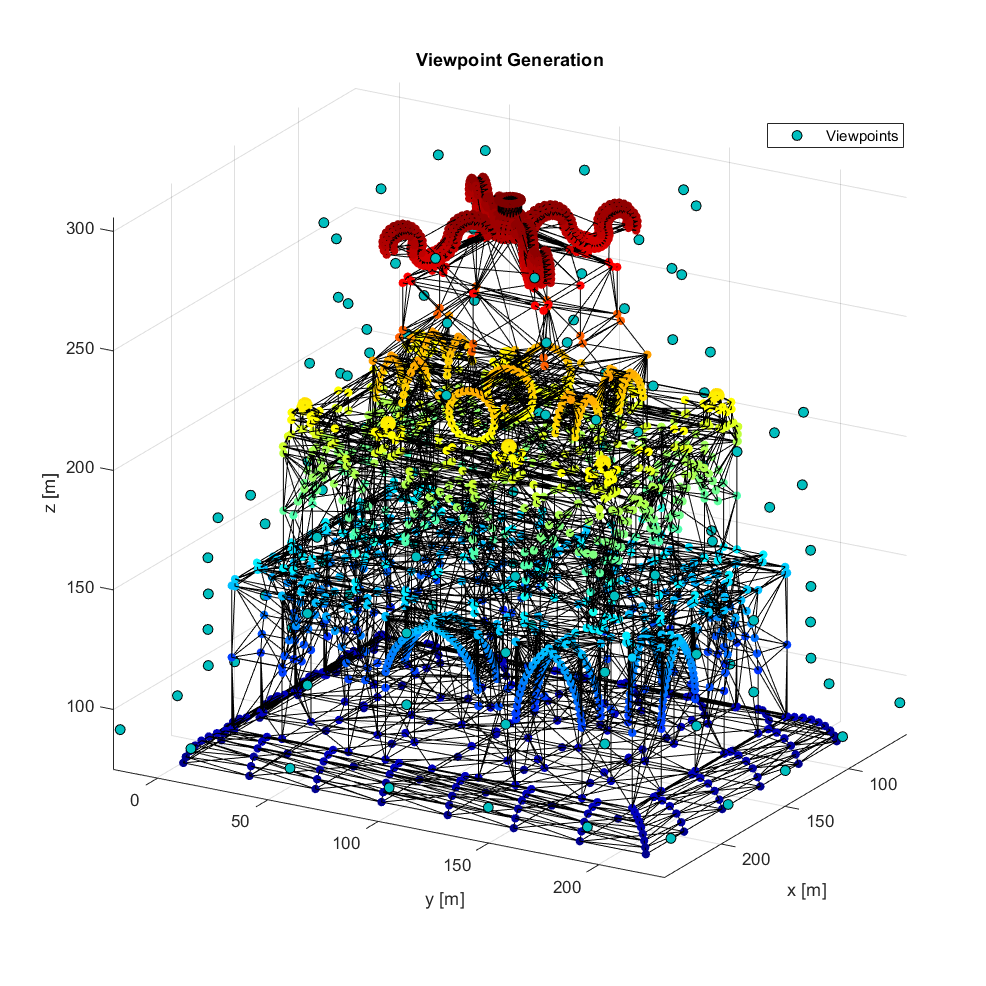}
         \caption{Generated viewpoints}
         \label{fig:gen}
     \end{subfigure}
     \begin{subfigure}[b]{0.38\textwidth}
         \centering
         \includegraphics[width=\textwidth]{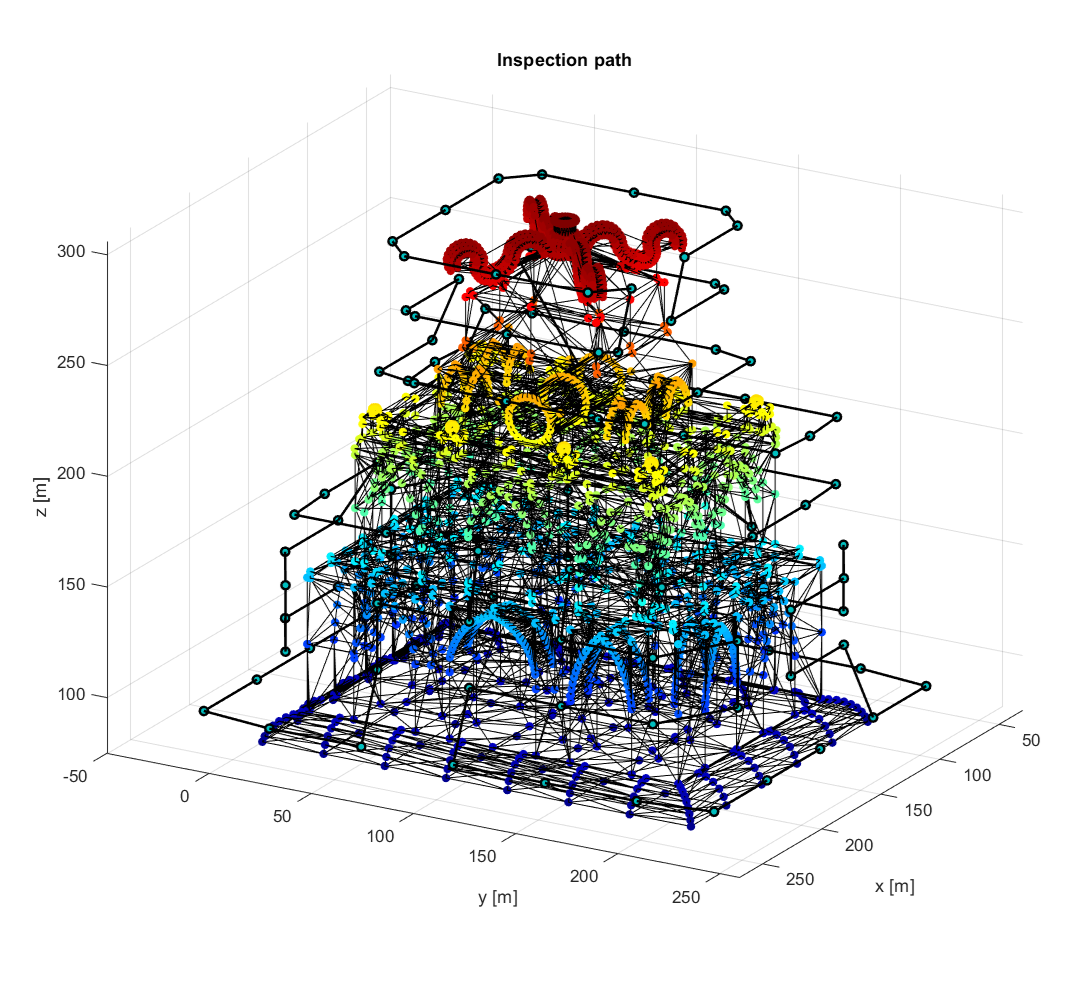}
         \caption{Inspection path}
         \label{fig:path_aco}
     \end{subfigure}
    \caption{Scenario 1 - The Turtle Tower}
    \label{fig:scen1}
\end{figure*}

\begin{figure*}
     \centering
     \begin{subfigure}[b]{0.325\textwidth}
         \centering
         \includegraphics[width=\textwidth]{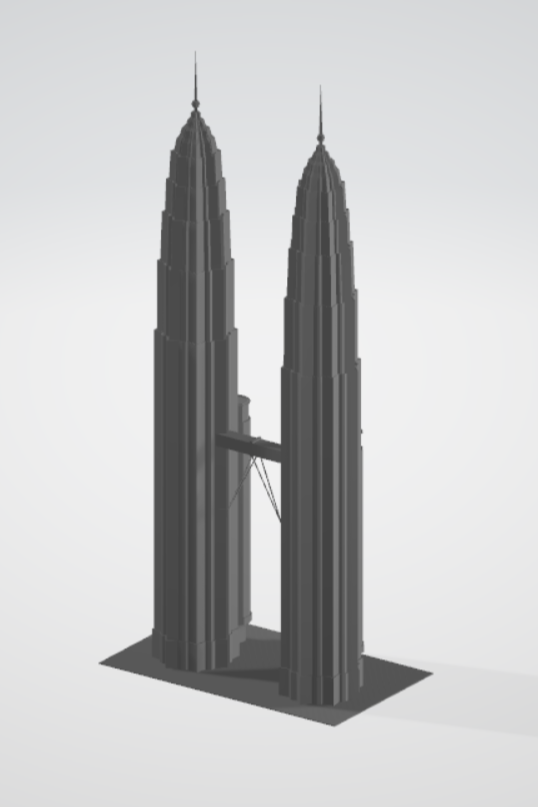}
         \caption{3D model}
         \label{fig:stl2}
     \end{subfigure}
     \begin{subfigure}[b]{0.325\textwidth}
         \centering
         \includegraphics[width=\textwidth]{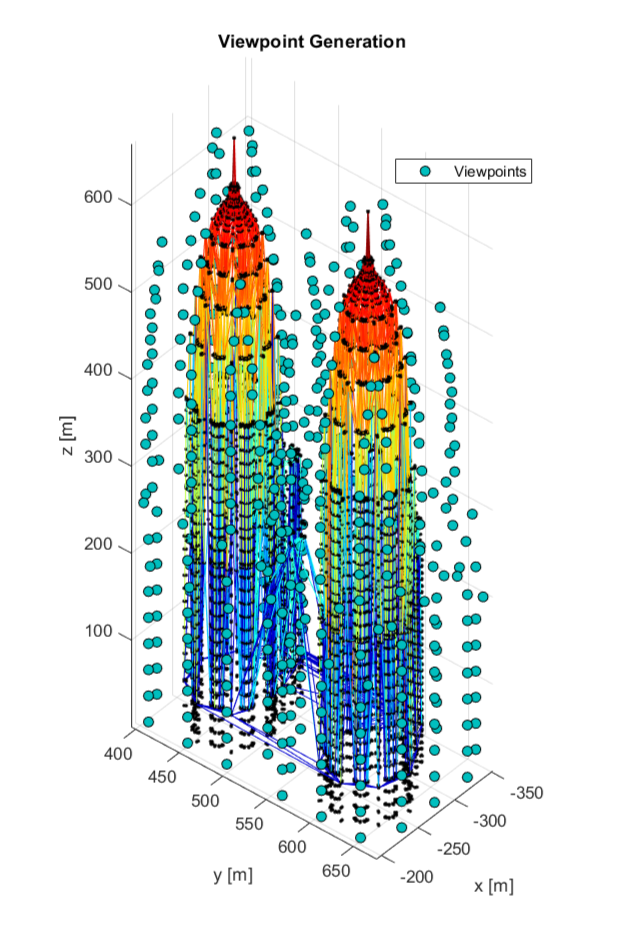}
         \caption{Generated viewpoints}
         \label{fig:gen2}
     \end{subfigure}
     \begin{subfigure}[b]{0.325\textwidth}
         \centering
         \includegraphics[width=\textwidth]{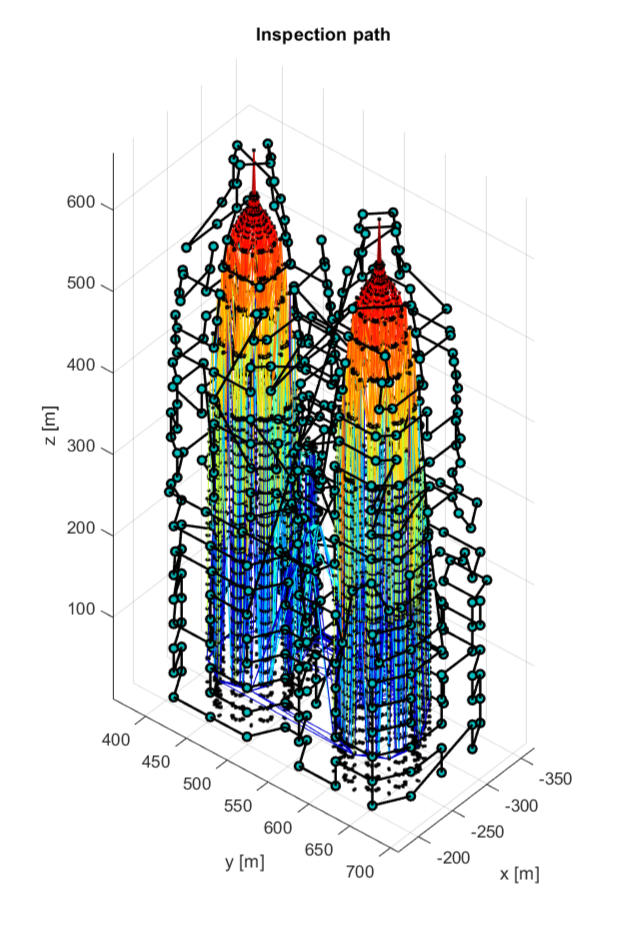}
         \caption{Inspection path}
         \label{fig:path_aco2}
     \end{subfigure}
    \caption{Scenario 2 - The twin tower}
    \label{fig:scen2}
\end{figure*}

\begin{figure*}
    \centering
    \begin{subfigure}[b]{0.4\textwidth}
     \centering
     \includegraphics[width=\textwidth]{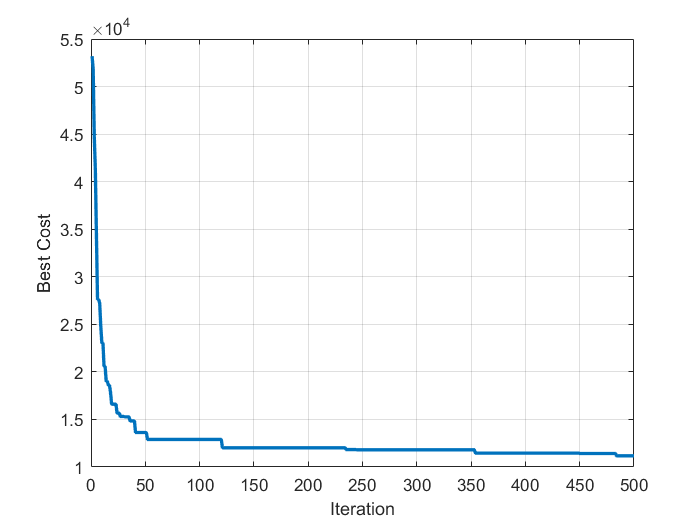}
     \caption{Scenario 1}
     \label{fig:cost}
    \end{subfigure}
    \begin{subfigure}[b]{0.4\textwidth}
     \centering
     \includegraphics[width=\textwidth]{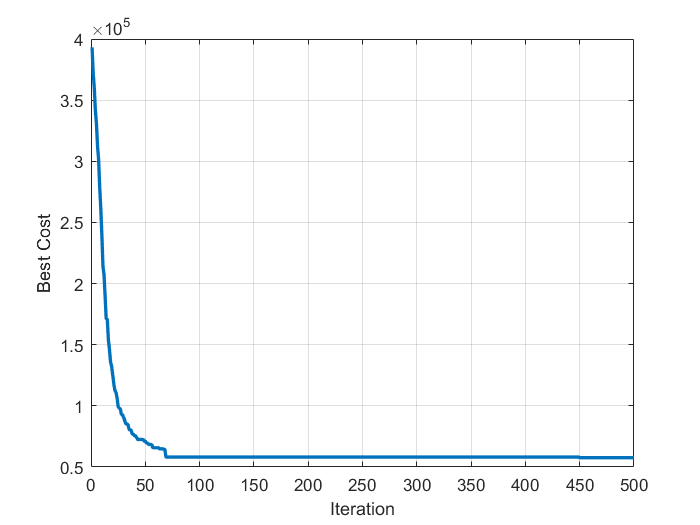}
     \caption{Scenario2}
     \label{fig:cost2}
    \end{subfigure}
    \caption{Best fitness values over iterations of the ACO algorithm}
    \label{fig:cost_aco}
\end{figure*}

In this section, simulation experiments are conducted to evaluate the performance of the proposed algorithm. 

\subsection{Scenario setup}
The UAVs used in this work are 3DR Solo drones that can be programmed to fly automatically via ground control station software. Each UAV is equipped with an IMX-100 camera for data collection, with the angles of view $\alpha =49.4^0$ and $\beta=63^0$. In the evaluation, we used 4 UAVs for the inspection, whose formation structure is depicted as Fig. \ref{fig:topology}. The distance between the infrastructure surface and inspection system is chosen as $d = 20$ m \cite{LI201483}. From \eqref{eqn:camera} and the image stitching requirement, the overall footprint of the formation is obtained as $w_f=48$ m and $h_f=34$ m.

Experiments have been carried out on two scaled real structure models described by STL data as shown in Fig. \ref{fig:stl} and Fig. \ref{fig:stl2}. The first model is the tower named Turtle Tower with a size of 150 m $\times$ 210 m $\times$ 231 m. The second on is a twin tower with a size of 119.14 m $\times$ 242.61 m $\times$ 669.75 m. The second structure is challenging to inspect as it includes two blocks connected by a bridge in the middle. In experiments, parameters of the ACO algorithm are chosen as $\alpha=\beta=1$, $\varrho=0.05$, $Q=1$, $w_1=1$, $w_2=2$. The number of ants is set to 100. The maximum iteration is set to 500.

\subsection{Path generation and IPP convergence}
Figures \ref{fig:gen} and \ref{fig:gen2} show the viewpoints generated to inspect the two structures. It can be seen that the viewpoints are well spread throughout the structures to cover all surfaces. Figures \ref{fig:path_aco} and \ref{fig:path_aco2} show the generated inspection paths. For the first structure, the path mainly goes along each layer from the bottom to the top similar to the simple back-and-forth strategy. The path for the second structure, however, is more complicated since the simple heuristic path generation strategies would not yield good results. The fitness values of the inspection paths during the optimization process are described in Fig \ref{fig:cost_aco}. It can be seen that they converge to small values in both scenarios and thus show the efficiency of the ACO algorithm in providing optimal solutions to the C-IPP problem.

\begin{table}
\centering
\caption{Fitness values of the paths generated by the BAF and ACO}
\label{tbl:plan}
\begin{tabular}{C{1cm}C{1.5cm}C{1.5cm}C{1.cm}C{1.5cm}} 
\hline
\multirow{2}{*}{Scenario} & \multirow{2}{*}{BAF} & \multicolumn{2}{c}{ACO} & \multirow{2}{*}{Improvement}  \\ 
\cline{3-4}
           &          & Mean       & Std        &                               \\ 
\hline
1                   & 1.2297e+04           & 1.1173e+04 & 54         & 9.14\%                   \\
\hline
2                   & 8.1639e+04           & 5.7582e+04 & 142        & 29.47\%                       \\
\hline
\end{tabular}
\end{table}

To further evaluate the performance of the proposed approach, a comparison with the back-and-forth (BAF) algorithm \cite{VAZQUEZCARMONA2022108125} is conducted in which the cost values of the inspection path generated by the BAF algorithm are obtained via the same fitness function designed in \eqref{eqn:J}. Table \ref{tbl:plan} shows the comparison results. It can be seen that for the first structure, the improvement of our algorithm is not significant due to the simplicity of the structure. For the second structure, however, the improvement of 29.47\%  clearly shows the superior performance of our algorithm compared to the BAF. It proves that the proposed method is capable of providing efficient inspection paths for 3D structures with different complexity.

%% file: root.bbl
\begin{thebibliography}{10}
\providecommand{\url}[1]{#1}
\csname url@samestyle\endcsname
\providecommand{\newblock}{\relax}
\providecommand{\bibinfo}[2]{#2}
\providecommand{\BIBentrySTDinterwordspacing}{\spaceskip=0pt\relax}
\providecommand{\BIBentryALTinterwordstretchfactor}{4}
\providecommand{\BIBentryALTinterwordspacing}{\spaceskip=\fontdimen2\font plus
\BIBentryALTinterwordstretchfactor\fontdimen3\font minus
  \fontdimen4\font\relax}
\providecommand{\BIBforeignlanguage}[2]{{%
\expandafter\ifx\csname l@#1\endcsname\relax
\typeout{** WARNING: IEEEtran.bst: No hyphenation pattern has been}%
\typeout{** loaded for the language `#1'. Using the pattern for}%
\typeout{** the default language instead.}%
\else
\language=\csname l@#1\endcsname
\fi
#2}}
\providecommand{\BIBdecl}{\relax}
\BIBdecl

\bibitem{PHUNG201725}
M.~D. Phung, C.~H. Quach, T.~H. Dinh, and Q.~Ha, ``Enhanced discrete particle
  swarm optimization path planning for {UAV} vision-based surface inspection,''
  \emph{Automation in Construction}, vol.~81, pp. 25--33, 2017.

\bibitem{8593930}
V.~Hoang, M.~Phung, T.~Dinh, and Q.~Ha, ``Angle-encoded swarm optimization for
  {UAV} formation path planning,'' in \emph{2018 IEEE/RSJ International
  Conference on Intelligent Robots and Systems (IROS)}, Oct 2018, pp.
  5239--5244.

\bibitem{WANG202213}
Y.~Wang, Y.~Li, F.~Yin, W.~Wang, H.~Sun, J.~Li, and K.~Zhang, ``An intelligent
  {UAV} path planning optimization method for monitoring the risk of unattended
  offshore oil platforms,'' \emph{Process Safety and Environmental Protection},
  vol. 160, pp. 13--24, 2022.

\bibitem{Wu2021}
Y.~Wu, S.~Wu, and X.~Hu, ``Multi-constrained cooperative path planning of
  multiple drones for persistent surveillance in urban environments,''
  \emph{Complex {\&} Intelligent Systems}, vol.~7, no.~3, pp. 1633--1647, Jun
  2021.

\bibitem{IVIC2023104709}
S.~Ivić, B.~Crnković, L.~Grbčić, and L.~Matleković, ``Multi-{UAV}
  trajectory planning for {3D} visual inspection of complex structures,''
  \emph{Automation in Construction}, vol. 147, p. 104709, 2023.

\bibitem{LI201483}
G.~Li, S.~He, Y.~Ju, and K.~Du, ``Long-distance precision inspection method for
  bridge cracks with image processing,'' \emph{Automation in Construction},
  vol.~41, pp. 83--95, 2014.

\bibitem{MANSOURI2018118}
S.~S. Mansouri, C.~Kanellakis, E.~Fresk, D.~Kominiak, and G.~Nikolakopoulos,
  ``Cooperative coverage path planning for visual inspection,'' \emph{Control
  Engineering Practice}, vol.~74, pp. 118--131, 2018.

\bibitem{doi:10.1080/17508975.2019.1613219}
R.~Panchalingam and K.~C. Chan, ``A state-of-the-art review on artificial
  intelligence for smart buildings,'' \emph{Intelligent Buildings
  International}, vol.~13, no.~4, pp. 203--226, 2021.

\bibitem{9341089}
W.~Jing, D.~Deng, Y.~Wu, and K.~Shimada, ``Multi-{UAV} coverage path planning
  for the inspection of large and complex structures,'' in \emph{2020 IEEE/RSJ
  International Conference on Intelligent Robots and Systems (IROS)}, Oct 2020,
  pp. 1480--1486.

\bibitem{SHANG2020113535}
Z.~Shang, J.~Bradley, and Z.~Shen, ``A co-optimal coverage path planning method
  for aerial scanning of complex structures,'' \emph{Expert Systems with
  Applications}, vol. 158, p. 113535, 2020.

\bibitem{Nagasawa2021}
R.~Nagasawa, E.~Mas, L.~Moya, and S.~Koshimura, ``Model-based analysis of
  multi-{UAV} path planning for surveying postdisaster building damage,''
  \emph{Scientific Reports}, vol.~11, no.~1, p. 18588, Sep 2021.

\bibitem{SON2015172}
H.~Son, F.~Bosché, and C.~Kim, ``As-built data acquisition and its use in
  production monitoring and automated layout of civil infrastructure: A
  survey,'' \emph{Advanced Engineering Informatics}, vol.~29, no.~2, pp.
  172--183, 2015, infrastructure Computer Vision.

\bibitem{BOLOURIAN2020103250}
N.~Bolourian and A.~Hammad, ``Lidar-equipped {UAV} path planning considering
  potential locations of defects for bridge inspection,'' \emph{Automation in
  Construction}, vol. 117, p. 103250, 2020.

\bibitem{9496675}
L.~V. Santana, A.~S. Brandão, and M.~Sarcinelli-Filho, ``On the design of
  outdoor leader-follower {UAV}-formation controllers from a practical point of
  view,'' \emph{IEEE Access}, vol.~9, pp. 107\,493--107\,501, 2021.

\bibitem{10.5555/3001460.3001507}
M.~Ester, H.-P. Kriegel, J.~Sander, and X.~Xu, ``A density-based algorithm for
  discovering clusters in large spatial databases with noise,'' in
  \emph{Proceedings of the Second International Conference on Knowledge
  Discovery and Data Mining}, ser. KDD'96.\hskip 1em plus 0.5em minus
  0.4em\relax AAAI Press, 1996, p. 226–231.

\bibitem{LIU2007240}
X.-Z. Liu, J.-H. Yong, G.-Q. Zheng, and J.-G. Sun, ``An offset algorithm for
  polyline curves,'' \emph{Computers in Industry}, vol.~58, no.~3, pp.
  240--254, 2007.

\bibitem{10.1007/978-981-13-6001-5_32}
K.~Chaudhari and A.~Thakkar, ``Travelling salesman problem: An empirical
  comparison between {ACO}, {PSO}, {ABC}, {FA} and {GA},'' in \emph{Emerging
  Research in Computing, Information, Communication and Applications}, N.~R.
  Shetty, L.~M. Patnaik, H.~C. Nagaraj, P.~N. Hamsavath, and N.~Nalini,
  Eds.\hskip 1em plus 0.5em minus 0.4em\relax Singapore: Springer Singapore,
  2019, pp. 397--405.

\bibitem{4129846}
M.~Dorigo, M.~Birattari, and T.~Stutzle, ``Ant colony optimization,''
  \emph{IEEE Computational Intelligence Magazine}, vol.~1, no.~4, pp. 28--39,
  Nov 2006.

\bibitem{VAZQUEZCARMONA2022108125}
E.~V. Vazquez-Carmona, J.~I. Vasquez-Gomez, J.~C. Herrera-Lozada, and
  M.~Antonio-Cruz, ``Coverage path planning for spraying drones,''
  \emph{Computers \& Industrial Engineering}, vol. 168, p. 108125, 2022.

\end{thebibliography}
